\documentstyle[epsfig]{aipproc}

\begin{document}
\title{Dileptons from P-Nucleus Collisions}

\author{Jen-Chieh Peng}
\address{Physics Division, Los Alamos National Laboratory\\
Los Alamos, New Mexico 87545}

\maketitle

\begin{abstract}
Recent results from fixed-target dimuon production experiments 
at Fermilab are presented. Various QCD tests using
dilepton production data are discussed. We emphasize that clear evidence
for scaling violation in the Drell-Yan process remains to be established.
Further theoretical and experimental work are needed to understand the
polarization of the Drell-Yan pairs.
We also discuss the nuclear medium effects for dilepton productions.
In particular, we discuss the use of Drell-Yan
data to deduce the energy-loss of partons traversing nuclear medium. 
\end{abstract}

\section*{Introduction}

The experimental detection of high-mass lepton pairs produced 
in hadronic reactions
has a long and rich history. The famous quarkonium 
states that revealed the existence of the charm and beauty quarks in 
the 1970s were discovered through their dilepton decay branches. They are
superimposed on a continuum, which was anticipated theoretically in 
1970~\cite{drell}, and is now known as the Drell-Yan (DY) 
process. The DY process, electromagnetic 
quark-antiquark annihilation, is closely related to the
deeply inelastic lepton scattering (DIS). By 1980, DY production 
was already a source of information about antiquark structure of the nucleon. 
Additionally, DY production with beams of pions and kaons yielded
the structure functions of these unstable particles for the first time.
Also notable in the history of the DY process were the discoveries of the 
$W^\pm$ and $Z^0$ particles in 1983, produced by a      
generalized (vector boson exchange) quark-antiquark annihilation mechanism.

New experimental work has been carried out in recent 
years by few but prolific collaborations working in the fixed-target 
programs at the CERN SPS accelerator and at Fermilab.
A series of fixed-target dimuon production experiments (E772, E789, E866)
have been carried out at Fermilab in the last 10 years. 
Some of the highlights from these experiments, namely the observation of
pronounced flavor asymmetry in the nucleon sea and the absence of antiquark
enhancement in heavy nuclei, are discussed by Garvey at this Conference.
In this paper, we will focus on other areas of dilepton physics studied in 
these experiments. To follow the main theme of this Conference, we first 
discuss the status of various QCD tests using dilepton productions. We
will then discuss the nuclear medium effects for dilepton productions. 
In particular, the relevance of the Fermilab experiments on the issue of
parton energy loss in nuclei will be presented.

\begin{figure}
\centerline{\epsfig{file=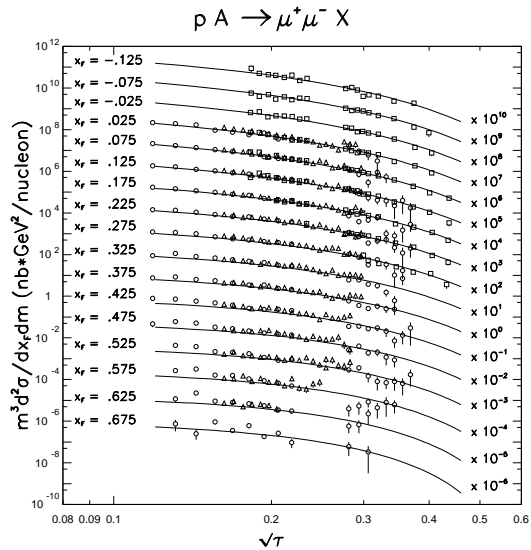,height=3.0in,width=3.0in}}
\vspace{10pt}
\caption{Proton-induced Drell-Yan production from experiments 
NA3 [2] (triangles) at 
400 GeV/c, E605 [3] (squares) at 800 GeV/c, and 
E772 [4] (circles) at 800 GeV/c. The lines are absolute (no 
arbitrary normalization factor)
next-to-leading order calculations for 
$p + d$ collisions at 800 GeV/c using the                  
CTEQ4M structure functions [5].}
\label{figsum}                                                     
\end{figure}

\section*{QCD tests in Dilepton Production}

In the ``Naive" Drell-Yan model,
the differential cross section is given as
\begin{eqnarray}
M^3 {d^2\sigma\over dM dx_F} = {8\pi\alpha^2\over 9}
{x_1 x_2\over x_1 + x_2} \times \sum_a e_a^2[q_a(x_1)
\bar q_a(x_2)+\bar q_a(x_1)q_a(x_2)]. \label{eq:dy}
\end{eqnarray}
Here $q_a(x)$ are the quark or antiquark structure functions of 
the two colliding hadrons evaluated at momentum fractions $x_1$ and 
$x_2$. The sum is over quark flavors. The Feynman-$x$ ($x_F$) is equal to
$x_1 - x_2$.

Although the simple parton model originally proposed by Drell and Yan
enjoyed considerable success in explaining many features of the
early data, it was soon realized that QCD corrections to the parton model
were required. Historically, two 
experimental features demanded theoretical improvement: first,
the experimental cross section was about a factor of 
two larger than the parton-model value, and second, 
the distribution of dilepton transverse momenta extended to much larger 
values than are characteristic of the convolution of intrinsic parton 
momenta. We now discuss several consequences of QCD corrections
to the DY observables.

\subsection*{Absolute Cross Sections and $p_T$ Distribution}

The inclusion of the NLO diagrams for the DY process brings excellent
agreements between the calculations and the data.
As an example, Figure~\ref{figsum} shows the NA3 data~\cite{na3} at 
400 GeV, together with the E605~\cite{e605} and E772~\cite{e772a} data at 800 GeV. 
The solid curve in Figure~\ref{figsum} corresponds to NLO calculation 
for 800 GeV $p+d$ ($\sqrt s$ = 38.9 GeV) and it describes
the NA3/E605/E772 data well. 

\begin{figure}
\centerline{\epsfig{file=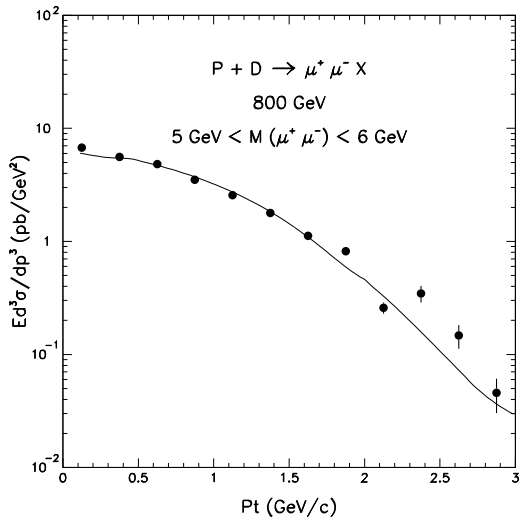,height=3.0in,width=3.0in}}
\vspace{10pt}
\caption{Comparison between the NLO calculation from Ref. [6]
and the E772 data [4].}
\label{e772pt1}                                                     
\end{figure}

Berger et al.~\cite{berger98} recently compared their NLO 
calculations with the E772 data.
As shown in Figure~\ref{e772pt1}, the $p_T$ distribution is well reproduced
by the calculation. At $p_T > 2$ GeV/c, the DY cross section is shown to be
dominated by processes involving gluons. This suggests the interesting 
possibility of probing gluon density using large $p_T$ DY events.

\subsection*{Scaling Violation}

The right-hand side of Eq. (1) is only a function of $x_1, x_2$ and is 
independent of the beam energies. This scaling property no longer holds
when QCD corrections to the DY are taken into account.

While logarithmic scaling violation is well established in 
DIS experiments,
it is not well confirmed in DY experiments at all. 
No evidence for
scaling violation is seen. As discussed in a recent review~\cite{plm}, 
there are mainly two reasons for this. First, unlike the DIS, 
the DY cross section is a convolution of two structure functions. Scaling 
violation implies that the structure functions rise for $x \leq 0.1$ and 
drop for $x \geq 0.1$ as $Q^2$ increases. For proton-induced
DY, one often involves a beam quark with $x_1 > 0.1$ and a target antiquark
with $x_2 < 0.1$. Hence the effects of scaling violation are partially 
cancelled. Second, unlike the DIS, the DY experiment can only probe 
relatively large $Q^2$, namely, $Q^2 > 16$ GeV$^2$ for a mass cut of 4 GeV. 
This makes it more difficult to observe the logarithmic variation of the 
structure functions in DY experiments.

\begin{figure}
\centerline{\epsfig{file=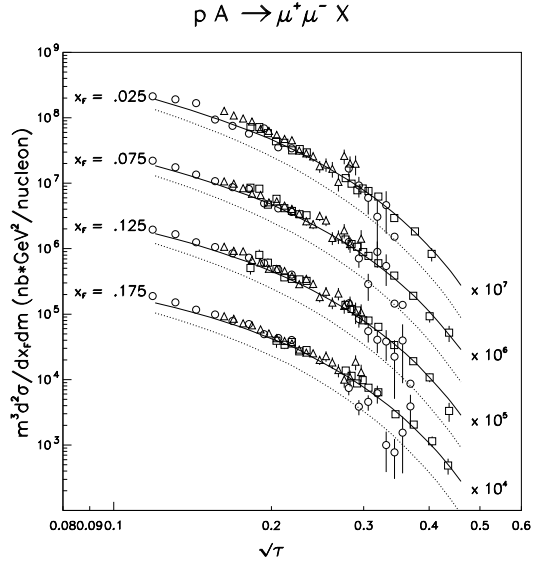,height=3.0in,width=3.0in}}
\vspace{10pt}
\caption{Comparison of DY cross section data with NLO
calculations using MRST [10] structure 
functions. Note that $\tau = x_1 x_2$.
The E772 [4], E605 [3], and
NA3 [2] data points are shown as circles, 
squares, and triangles, respectively.
The solid curve corresponds to fixed-target 
p+d collision at 800 GeV, while the
dotted curve is for p+d collision at $\sqrt s$ = 500 GeV.}
\label{rhicfig1}
\end{figure}

Possible indications of scaling violation in DY process have been reported
in two pion-induced experiments, E326~\cite{e326} at 
Fermilab and NA10~\cite{na10} at CERN.
E326 collaboration compared their 225 GeV $\pi^- + W$ DY cross 
sections against calculations with and without scaling violation.
They observed better agreement when scaling violation is included.
This analysis is subject to the uncertainties associated with 
the pion structure functions, as well as the nuclear effects of the $W$ target.
The NA10 collaboration measured $\pi^- + W$ DY cross sections at three
beam energies, namely, 140, 194, and 286 GeV. By checking the ratios of the 
cross sections at three different energies, NA10 largely avoids the 
uncertainty of the pion structure functions. However, the relatively small 
span in $\sqrt s$, together with the complication of nuclear effects, make 
the NA10 result less than conclusive.

RHIC provides an interesting opportunity for unambiguously establishing 
scaling violation in the DY process. 
Figure~\ref{rhicfig1} shows the predictions for $p+d$ at $\sqrt s$ = 500 GeV. 
The scaling-violation accounts for a factor of two
drop in the DY cross sections when $\sqrt s$ is increased from 38.9 GeV
to 500 GeV. It appears quite feasible to establish scaling violation in
DY with future dilepton production experiments at RHIC.

\subsection*{Decay Anugular Distributions}

In the parton model, the angular distribution of dileptons is
characteristic of the decay of a transversely polarized virtual photon, 
\begin{eqnarray}
{d\sigma \over d\Omega} = \sigma_0 ( 1 + \lambda cos^2 \theta),
\label{eq:dytheta}
\end{eqnarray}
where $\theta$ is the polar angle of the lepton in the virtual photon
rest frame and $\lambda = 1$. Early experimental 
data from both 
pion and proton beams~\cite{kenyon} were consistent with this form
but had large statistical errors.

\begin{figure}
\centerline{\epsfig{file=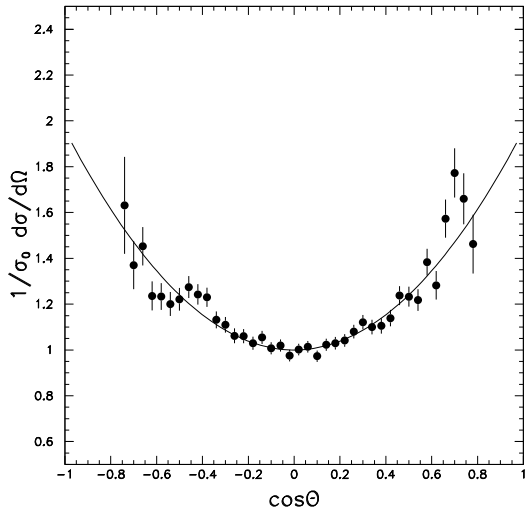,height=3.0in,width=3.0in}}
\vspace{10pt}
\caption{Drell-Yan angular distribution from Fermilab E772 [12]:
$p + Cu$ collisions at 800 GeV/c. The dimuons cover the mass region 
$11\leq M_{\mu^+\mu^-} \leq 17$ GeV/c$^2$ with 
$-0.3\leq x_F\leq 0.8$ and $p_t \leq$ 6 GeV/c. Mean
values for $p_t$, $x_F$, and $M$ are 1.4 GeV/c, 0.16, and
11.9 GeV/c$^2$, respectively. The solid curve is a fit to the data
with the form $1 + \lambda cos^2\theta$, where $\lambda$ is 
$0.96 \pm .04 \pm .06$.}
\label{dyang}
\end {figure}

Recently, E772 has performed a high-statistics study of the angular 
distribution for DY events~\cite{QM96} 
with masses above the $\Upsilon$ family
of resonances. About 50,000 events were recorded from 
800 GeV/c $p + Cu$ collisions, using a copper beam dump
as the target. Figure~\ref{dyang} shows the 
acceptance-corrected angular distribution, 
integrated over the kinematic variables.
Analyzed in the Collins-Soper reference frame~\cite{collins}, the data yield 
$\lambda = 0.96\pm 0.04\pm 0.06$(systematic).

Including higher-order QCD corrections to the DY process~\cite{oakes,collinsa}
results in the more complicated form of 
the angular distribution,
\begin{eqnarray}
{d\sigma \over d\Omega} \propto 1 + \lambda cos^2 \theta + 
\mu sin 2\theta cos \phi + {\nu \over 2} sin^2 \theta cos 2 \phi, 
\label{eq:dyangdist1}
\end{eqnarray}
where $\phi$ is the azimuthal angle and $\lambda$,
$\mu$, and $\nu$ are angle-independent parameters. NLO
calculations predict~\cite{chiappetta}
small deviations from $1 + cos^2 \theta$ ($\leq 5\%$) for $p_t$
below 3 GeV/c. The relevant scaling parameter for the magnitude
of these deviations is $p_t / Q$, implying that NLO 
corrections become important when $p_t \simeq Q$. 
A relation, $1 -\lambda -2\nu = 0$, developed by Lam \& Tung~\cite{lam},
is analogous to the Callan-Gross relation in DIS. Measurements with
pion beams at CERN~\cite{NA10angdist} and at Fermilab~\cite{E615a}
have shown that the Lam-Tung relation is clearly violated at large $p_t$.

Pion-induced DY experiments have unexpectedly shown that
transverse photon polarization changes to
longitudinal ($\lambda \simeq -1$) at 
large $x_F$~\cite{NA10angdist,E615a,NA3angdist,E615}.
The $x_F$ dependence of $\lambda$ is qualitatively consistent with 
a higher-twist model originally proposed by Berger \& 
Brodsky~\cite{berger,bergera}. However, the quantitative 
agreement is poor. The model's
basis can be described as follows.
As $x_F$ of the muon pair approaches unity, the Bjorken-$x$ (momentum
fraction) of the annihilating projectile parton must also be
near unity. Thus, the whole pion 
contributes to the DY process. This can be treated with perturbation 
theory, with the result that the transverse polarization of the
virtual photon becomes longitudinal. The angular distribution at
large $x_F$ becomes

\begin{eqnarray}
{d\sigma \over d\Omega} \propto (1-x)^2 (1 + \lambda cos^2 \theta) + 
\alpha sin^2 \theta, \label{eq:dyangdist12}
\end{eqnarray}

\noindent where $\alpha$ is $\propto$ $p_t^2 / Q^2$.

Eskola et al~\cite{eskola} have shown that an improved treatment
of the effects of nonasymptotic kinematics greatly improves
quantitative agreement with the $\lambda$ values from the pion data. 
Brandenburg et al~\cite{brandenburg} have extended the higher
twist model to specifically include pion bound-state effects.
They predict values for $\lambda$, $\mu$ and $\nu$ that are in
good agreement with the pion data at large $x_F$. Unfortunately,
the results are quite sensitive to the choice of the pion Fock
state wave functions, which are not well constrained by experimental
data.

\section*{Nuclear Medium Effects of Dilepton Production}

\begin{figure}
\centerline{\epsfig{file=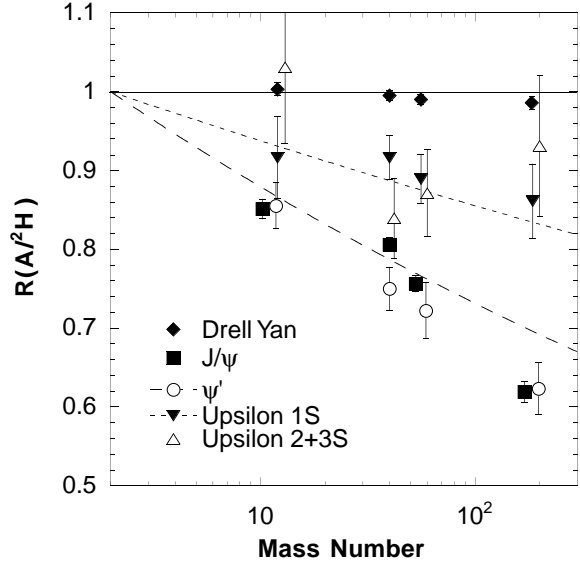,height=3.0in,width=3.0in}}
\vspace{10pt}
\caption{ Ratios of heavy-nucleus to deuterium
integrated yields per nucleon for 800 GeV proton production of dimuons from the 
Drell-Yan process and from decays of the $J/\psi$, $\psi '$, $\Upsilon 
(1S)$, and $\Upsilon (2S+3S)$ states [7]. The short 
dash and long dash 
curves represent the approximate nuclear dependences for the $b\bar b$ 
and $c\bar c$ states, $A^{0.96}$ and $A^{0.92}$, respectively.}
\label{rhicfig4}
\end{figure}

From a high-statistics measurement of dilepton production in 800 GeV
proton-nucleus interaction, the target-mass dependence of DY, $J/\Psi$,
$\Psi^\prime$, and $\Upsilon$ productions have 
been determined in E772~\cite{e772b,e772c,e772d}.
As shown in Figure~\ref{rhicfig4}, different 
nuclear dependences are observed for
different dilepton processes. While the DY process shows almost no nuclear
dependence, pronounced nuclear effects are seen for the production of
heavy quarkonium states. E772 found that $J/\Psi$ and $\Psi^\prime$
have similar nuclear dependence. The nuclear dependences for $\Upsilon$,
$\Upsilon^\prime$ and $\Upsilon^{\prime \prime}$ are less than that observed
for the $J/\Psi$ and $\Psi^\prime$. Within statistics, the various 
$\Upsilon$ resonances also have very similar nuclear dependences.

\begin{figure}
\centerline{\epsfig{file=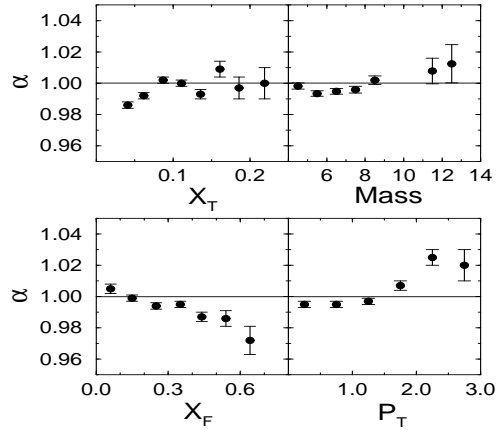,height=3.0in,width=3.0in}}
\vspace{10pt}
\caption{Nuclear dependence coefficient $\alpha$ for 800 GeV p+A
Drell-Yan process versus various kinematic variables [26].}
\label{rhicfig5}
\end{figure}

\begin{figure}
\centerline{\epsfig{file=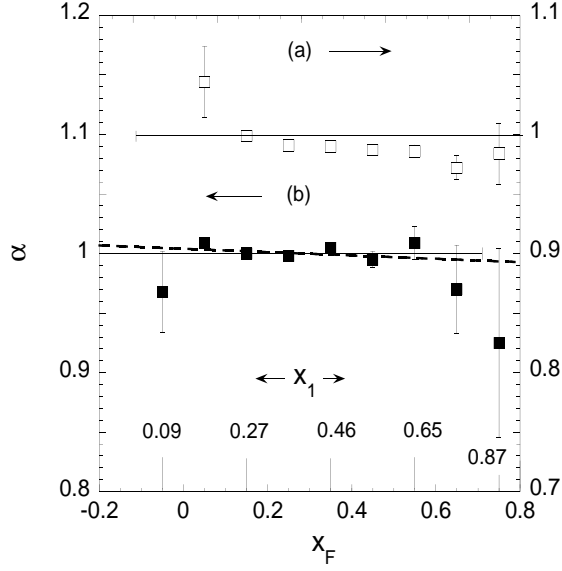,height=3.0in,width=3.0in}}
\vspace{10pt}
\caption{  Nuclear dependence coefficient $\alpha$ for the Drell-Yan 
process [26] versus
$x_F$ for (a) $x_2\leq 0.075$, right scale, and  (b) $x_2\geq 0.075$, left 
scale. The thin solid lines show $\alpha =1$. The dashed line is a 
linear least-squares fit to the lower points. Also shown is the 
mean value of $x_1$ for (b).}
\label{rhicfig6}
\end{figure}

\begin{figure}
\centerline{\epsfig{file=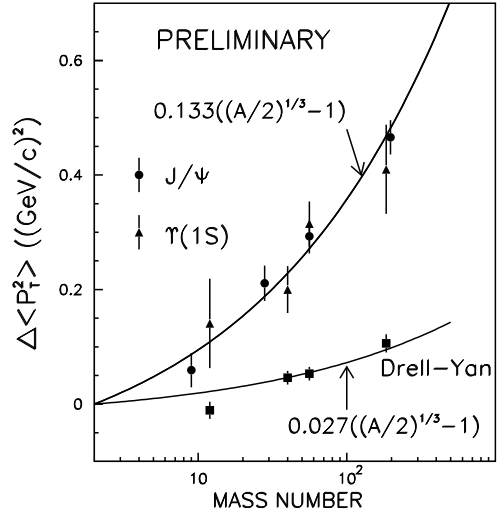,height=3.0in,width=3.0in}}
\vspace{10pt}
\caption{The change of mean $p_t^2$ for nuclear target, 
$\Delta \langle p_t^2 \rangle = \langle p_t^2 \rangle (A) -
\langle p_t^2 \rangle (D)$, for 800 GeV p+A Drell-Yan process and
$J/\Psi$ and $\Upsilon(1S)$ productions. The solid curves are best
fits to the A-dependence of $\Delta \langle p_t^2 \rangle$. The
$J/\Psi$ and $\Upsilon (1S)$ productions have identical 
curves for the A-dependence
fits.}
\label{rhicfig7}
\end{figure}

Although the integrated DY yields in E772 show little nuclear dependence,
it is instructive to examine the DY nuclear dependences on
various kinematic variables. Using the simple $A^\alpha$ expression 
to fit the DY nuclear dependence, the values of $\alpha$ are shown in 
Figure~\ref{rhicfig5} 
as a function of $x_T (x_2)$, $M$, $x_F$, and $p_t$.
Several features are observed:

\begin{enumerate}

\item A suppression of the DY yields from heavy nuclear targets is seen
at small $x_2$. This is consistent with the shadowing effect observed
in DIS. In fact, E772 provides the only experimental evidence for
shadowing in hadronic reactions. The reach of small $x_2$
in E772 is limited by the mass cut ($M \geq 4$ GeV) and by the
relatively small center-of-mass energy (recall that $x_1 x_2 = M^2/s$). p-A
collisions at RHIC clearly offer the exciting opportunity to extend the
study of shadowing to much smaller $x$.

\item $\alpha (x_F)$ shows an interesting trend, namely, it decreases
as $x_F$ increases. It is tempting to attribute this behavior to 
initial-state energy-loss effect. However, there is a strong correlation
between $x_F$ and $x_2$ ($x_F = x_1 -x_2$), and it is essential to
separate the $x_F$ energy-loss effect from the $x_2$ shadowing effect. 
Figure~\ref{rhicfig6} shows $\alpha$ versus $x_F$ 
for two bins of $x_2$, one in the
shadowing region ($x_2 \leq 0.075$) and one outside of it ($x_2 \geq 0.075$).
There is no discernible $x_F$ dependence for $\alpha$ once one stays outside
of the shadowing region. Therefore, the 
apparent suppression at large $x_F$ in Figure~\ref{rhicfig5}
reflects the shadowing effect at small $x_2$ rather than the energy-loss effect.

\item $\alpha (p_t)$ shows an enhancement at large $p_t$. This is reminiscent
of the Cronin Effect~\cite{cronin} where the broadening in $p_t$ distribution
is attributed to multiple parton-nucleon scatterings. It is instructive to 
compare the $p_t$ broadening for DY process and quarkonium production.
Figure~\ref{rhicfig7} shows $\Delta\langle p_t^2\rangle$, 
the difference of mean $p_t^2$ between
p-A and p-D interactions, as a function of A for DY, J/$\Psi$, 
and $\Upsilon(1S)$
productions at 800 GeV. The DY and $\Upsilon$ data 
are from E772~\cite{e772e}, while the
$J/\Psi$ results are from E789~\cite{e789}, E771~\cite{e771}, and
preliminary E866 analysis~\cite{leitch}. 
More details on this analysis will be 
presented elsewhere~\cite{e772e}. Figure~\ref{rhicfig7} shows 
that $\langle p_t^2\rangle$ is well described by
the simple expression $a + b A^{1/3}$. It also shows that the
$p_t$ broadening for $J/\Psi$ is very similar to $\Upsilon$,
but significantly larger (by a factor of 5) than the DY. A factor of
9/4 could be attributed to the color factor of the initial gluon in
the quarkonium production versus the quark in the DY process. 
The remaining difference
could come from the final-state multiple scattering effect which is absent
in the DY process.

\end{enumerate}
 
Baier et al.~\cite{baier1} have 
recently derived a relationship between the partonic 
energy-loss due to gluon bremsstrahlung and the mean $p_t^2$ broadening
accumulated via multiple parton-nucleon scattering:
\begin{eqnarray}
-dE/dz = {3\over 4}~ \alpha_s~ \Delta\langle p_t^2\rangle. \label{eq:dedx}
\end{eqnarray}
This non-intuitive result states that the total energy loss
is proportional to square of the path length traversed by the incident partons.
From Figure~\ref{rhicfig7} and 
Eq.~\ref{eq:dedx}, we deduce that the mean total energy loss,
$\Delta E$, for the p+W
DY process is $\approx 0.6$ GeV. Such an energy-loss is too small to
cause any discernible effect in the $x_F$ (or $x_1$) nuclear dependence.
As shown in Figure~\ref{rhicfig6}, the dashed curve corresponds 
to $\Delta E = 2.0 \pm 1.7$ GeV (for p+W), and the E772 data are consistent with
Eq.~\ref{eq:dedx}. A much more sensitive test for Eq.~\ref{eq:dedx} could 
be done at RHIC, where the energy-loss effect is expected to be much enhanced
in A-A collision~\cite{baier2}.

\end{document}